\documentclass[prl,twocolumn,preprintnumbers,amsmath,amssymb]{revtex4}

\usepackage{color}
\usepackage{graphicx}
\usepackage{dcolumn}
\usepackage{bm}

\begin{document}

\title{
Ab initio study of a mechanically gated molecule: From weak
to strong correlation }
\author{A. Greuling$^1$}
\author{M. Rohlfing$^1$}
\email{Michael.Rohlfing@uos.de}
\author{R. Temirov$^{2}$}
\author{F.S. Tautz$^{2}$}
\author{F.B. Anders$^3$}
\affiliation{
  $^1$Fachbereich Physik, Universit\"at Osnabr\"uck,
      Barbarastra{\ss}e 7, 49069 Osnabr\"uck, Germany\\
  $^2$Institut f\"ur Bio- and Nanosysteme, Forschungszentrum J\"ulich,
      52425 J\"ulich, Germany\\
  $^3$Fakult\"at f\"ur Physik,
      TU Dortmund, Otto-Hahn-Stra{\ss}e 4,
      44227 Dortmund, Germany
}

\date{\today}

\begin{abstract}

The electronic spectrum of a chemically contacted molecule
in the junction of a scanning tunneling microscope can be modified
by tip retraction. We analyze this effect by a combination of
density functional, many-body perturbation and numerical
renormalization group theory, taking into account both the
non-locality and the dynamics of electronic correlation. Our
findings, in particular the evolution from a broad quasiparticle
resonance below to a narrow Kondo resonance at the Fermi energy,
correspond to the experimental observations.

\end{abstract}


\maketitle

\begin{figure}
\scalebox{0.45}{\includegraphics{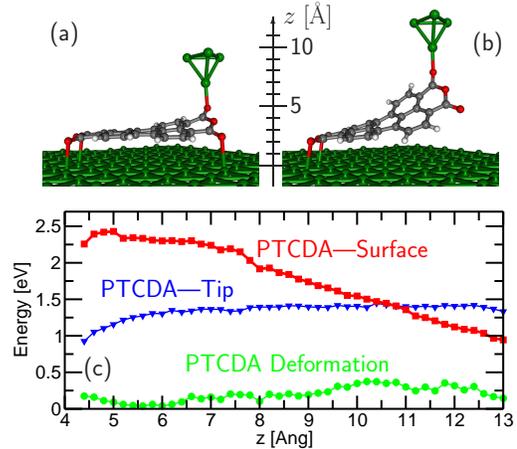}}
\caption{ \label{fig_DFT} 
DFT-LDA data of tip/PTCDA/Ag(111) junction.
(a, b) Two representative
configurations at tip-surface distances of $z$\,=\,7\,\AA \, and
$z$\,=\,10\,\AA, respectively. For better visibility only a
section of the tip and the substrate are shown. (c) Bond energies
of the PTCDA-tip and PTCDA-substrate bonds as a function of tip
height $z$. At each $z$ the molecule has been fully relaxed. The
internal deformation energy of the PTCDA reflects the bending and
twisting of its perylene core. }
\end{figure}

\begin{figure}
\scalebox{0.45}{\includegraphics{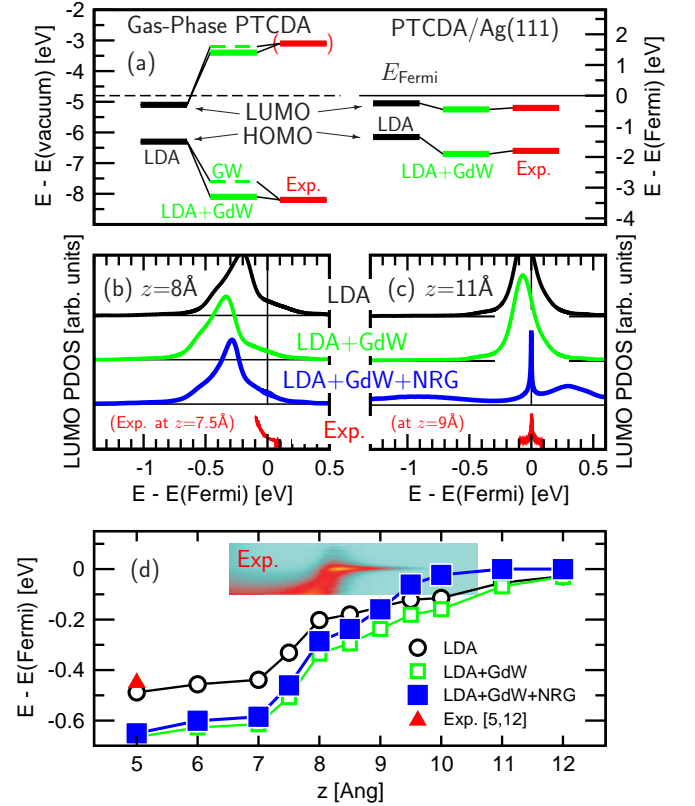}}
\caption{ 
\label{fig_LUMO} 
(a) HOMO and LUMO energies of gas-phase PTCDA and
PTCDA/Ag(111) 
(peak of the PDOS \cite{footnote0}; without STM tip). 
Experimental data are from
Ref.~\cite{Dori06,footnote3} (gas phase) and 
Ref.~\cite{Kraft06} (on Ag(111)).
%
%
%
The two energy scales differ by $E_F = E_{\rm vac} - E_W$ 
($E_W$ = work function of Ag(111)). (b,c)
LUMO spectra for the tip/PTCDA/Ag(111) junction \cite{footnote0} 
at tip heights of $z$\,=\,8\,\AA \, and $z$\,=\,11\,\AA.
The experimental spectra are cuts through the dI/dV map of panel d,
at $z$-values of 7.5 \AA\, and 9 \AA.
(d) LUMO peak positions and experimental dI/dV color map, ranging 
from 0\,$G_0$ (light blue) to 0.10\,$G_0$ (yellow).
Experiments were performed by approach-retraction cycles of
the tip, with bias voltages increasing in steps of 1\,mV.
$dI/dV$ was detected with a lock-in amplifier, modulation
amplitude 4\,mV and modulation frequency 2.9\,kHz.
For bias voltages exceeding $\pm$100 meV the tip-PTCDA contact
becomes unstable.
The experimental data point at --0.45 eV is an estimate from
Ref. \onlinecite{Temirov08} (--0.35 eV with the tip attached to the
PTCDA monolayer) and Ref. \onlinecite{Kilian08} (difference --0.1 eV 
between single-molecule and monolayer spectra).
}
\end{figure}

Both the control of the geometric structure of a molecular
junction and the systematic manipulation of its electronic
structure by an external parameter are of central
importance for molecular electronics
\cite{Lafferentz09}. So far, the best control over the geometric
junction structure is reached in experiments based on scanning
tunnelling microscopy (STM) \cite{Fernandez08,Wang09}, because
they allow the selection of an individual molecule in a specific
environment and its contacting with the STM tip at a defined
position within the molecule (e.g.~ref~\cite{BerndtPRL}). On the
other hand, the most common approach to tuning the electronic
structure, electrical gating, is difficult to combine with STM.
However, in a recent experimental study \cite{Temirov08} we
contacted a surface-adsorbed molecule with an STM tip and peeled
it off the surface by tip retraction, as shown schematically in
Fig.~\ref{fig_DFT}. Spectroscopic data recorded during tip
retraction revealed a \emph{mechanical} gating effect, in the
sense that one of the molecular orbitals responds to the
structural change and shifts with respect to the Fermi level
($E_F$) of the substrate, before becoming pinned at $E_F$
(Fig.~\ref{fig_LUMO}d).

In this letter we use this experiment, which was carried out on
3,4,9,10-perylene-tetracarboxylic-dianhydride (PTCDA) adsorbed on
the Ag(111) surface, as a motivation for a theoretical study of the
interplay between the geometric structure and the electronic
spectrum of a molecular junction. 
The mechanical gating is subject to subtle details of electronic
correlation, e.g.  screening of the intramolecular Coulomb repulsion by 
the electrodes, which is not described by conventional 
density-functional theory (DFT). 
We therefore propose the following strategy: DFT addresses the
atomistic details of the junction structure but does not provide
reliable, well-founded spectral data. To evaluate the electronic
spectrum, we combine DFT with 
many-body perturbation theory (MBPT) \cite{Onida02} for non-local 
correlation and with numerical renormalization-group (NRG) theory
\cite{BullaCostiPruschke2008} for
correlation dynamics beyond the mean-field level. We stress that
only by combining DFT, MBPT and NRG a well-founded picture emerges, 
which can by systematically compared to the data presented in
Ref.~\cite{Temirov08} and the (more comprehensive) experiment in
Fig.~\ref{fig_LUMO}d.

We first discuss the geometric structure of the tip/PTCDA/Ag(111)
junction in the framework of the local density approximation (LDA)
to DFT \cite{footnote1}. PTCDA molecules adsorb on the Ag(111)
surface in a flat-lying configuration
\cite{Hauschild05,Du06,Kraft06,Kilian08,Rohlfing08}. Several DFT
studies of the adsorption have been published (for details, see
Ref.~\cite{Rohlfing07} and references therein). Following
experiment \cite{Temirov08}, we place the tip, which in our
simulation consists of ten Ag atoms in a pyramidal shape
\cite{footnote2}, above one of the carboxylic oxygen atoms and
approach it to the surface. At a tip-surface distance of
$z$\,=\,6.2\,\AA \, (6.7\,\AA\, in experiment \cite{Temirov08})
the oxygen atom jumps up and forms a covalent bond with the tip
apex atom. The tip can then be moved up and down reversibly,
forcing the oxygen atom and the attached section of the PTCDA
molecule to follow, resulting in a peeling-like motion
(Fig.~\ref{fig_DFT}a-b). Throughout the paper, $z$ specifies the
vertical distance between the tip apex atom and the uppermost
surface layer.

Some representative DFT-LDA data of the structure and energetics of the
junction are displayed in
Fig.~\ref{fig_DFT}. Fig.~\ref{fig_DFT} a and b shows two typical
configurations. At a tip-surface distance of $z$\,=\,7\,\AA \,
only a small section of the molecule is detached from the surface,
while at $z$\,=\,10\,\AA \, about half of the molecule has lost
contact with the substrate. The continuous nature of the peeling
process as a function of $z$ is reflected in the smooth behavior
of the molecule-tip and molecule-surface bond energies
(Fig.~\ref{fig_DFT}c) and in a smooth $z$-dependence of all
intramolecular structural parameters.

The most interesting feature of the electronic structure of the
junction is the PTCDA LUMO (lowest unoccupied molecular orbital),
because this is the orbital for which the gating effect as a
function of the external parameter $z$ is observed
\cite{Temirov08}. In the gas phase, the LUMO is found
$\sim$\,2\,eV above the Ag(111) Fermi level 
(Fig.~\ref{fig_LUMO}a). Upon adsorption, the LUMO is lowered in
energy below the Ag(111) Fermi level due to the metallic
polarizability of the substrate
\cite{Neaton06,Thygesen09,Rohlfing10} and thus becomes partially
occupied with about 1.8 electrons
(counterbalanced by back-donation from other molecular 
states to the surface).

The partial occupation of the LUMO (which changes as
a function of $z$; see below) causes the high sensitivity of the
LUMO spectrum to the junction structure: 
In experiment, for $z$ below 8.2 \AA\ the LUMO approaches the Fermi
energy at a steep slope of $\sim$0.2-0.3 eV/\AA\ and turns into a
sharp resonance as soon as it reaches $E_F$ at $z$=8.2 \AA. At
$z=$\,9.7\,\AA, the FWHM of the LUMO peak is 14 meV. We note that
although a mechanical gating effect is observed already in DFT-LDA
(top curves in Fig.~\ref{fig_LUMO}b-c and open circles in
Fig.~\ref{fig_LUMO}d), there is a large discrepancy with the
experimental data: The DFT-LDA LUMO reaches the Fermi level far
too late (at 12 \AA), and it does not sharpen as much as in
experiment. Because of this sharpening, the experimental resonance
at the Fermi level has been discussed in terms of dynamic
correlations (Kondo effect) \cite{Temirov08}.

A systematic approach to the Kondo effect is achieved by taking the 
single-particle
mean-field spectrum of the LUMO (at each $z$) as a starting point for 
a NRG simulation (see below for details).
However, the DFT-LDA spectrum should not be used for this purpose since
it lacks physical significance, being essentially an auxiliary
quantity within a ground-state total-energy approach.
A proper single-particle mean-field
spectrum to be used as input to NRG must refer to electronic
excitations (i.e., removal or addition of one electron)
\cite{Onida02}.
In particular, these excitations are subject to non-local correlation
effects that are not included in DFT-LDA.

Mean-field spectra
which take non-local correlation into account can be addressed
within the standard $GW$ approximation of MBPT \cite{Onida02}. In
the present case, however, a full $GW$ calculation is too
demanding. We therefore employ a simplified, perturbative
LDA+$GdW$ approach which yields reliable QP energies
$E^{GdW}_n := E^{\rm LDA}_n+ \Delta^{GdW}_n$ 
on top of DFT-LDA from a QP
Hamiltonian \cite{Gygi89,Fiorentini95,Rohlfing10}
\begin{equation}
\label{eq_1} \hat{H}^{{\rm QP, LDA}+GdW} := \hat{H}^{\rm LDA}  + i G
(W-W_{\rm metal}) \quad ,
\end{equation}
in which the QP self-energy correction $\Delta^{GdW} := i G (W-W_{\rm
metal})$ results from the difference between the correctly
screened Coulomb interaction ($W$) and an interaction $W_{\rm
metal}$ from hypothetical metallic screening (see
Ref.~\cite{Rohlfing10} for details).
The resulting QP corrections to DFT-LDA differ from 
standard $GW$ QP corrections by less than 20 \%
\cite{Wang83,Gygi89,Fiorentini95,Rohlfing10}.
For gas-phase PTCDA, LDA+$GdW$ can easily be checked against a
standard $GW$ calculation and against the experiment.
Fig.~\ref{fig_LUMO}a shows that the $GdW$ shifts
($\Delta^{GdW}$\,=\,$-$1.8\,eV for the HOMO and +1.7\,eV for the
LUMO) agree well with standard $GW$ results (both with our
own and with those of Ref.~\cite{Dori06}), in particular for the
LUMO, and with experiment \cite{Dori06,footnote3}, establishing
LDA+$GdW$ as a suitable method.

\begin{figure}
\scalebox{0.50}{\includegraphics{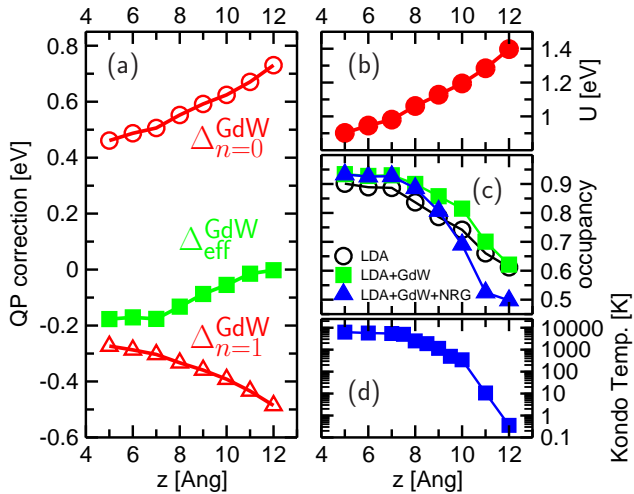}}
\caption{ \label{fig_Parameters} Properties of the PTCDA LUMO
at varying tip-surface distances $z$. (a) Mean-field QP
correction (from LDA+$GdW$) of the LUMO for various
occupancies, i.e.~(from top to bottom) for the state being empty,
being partially occupied (at $n_{{\sf LDA}+GdW}$), and being filled.
(b) Intra-state interaction $U$ (from Eq.~(\ref{eq_3})). (c)
Occupancy of the LUMO, resulting from LDA and LDA+$GdW$. Note
that the number of electrons in the state is 2$\cdot$$n$. 
(d) Resulting Kondo temperature.
}
\end{figure}
When the molecule is adsorbed on the surface, three features show
up in $\Delta^{GdW}$ and $E^{GdW}$:
(i) Due to the metallic nature of the substrate,
which screens the interaction inside the molecule non-locally
\cite{Neaton06,Thygesen09,Rohlfing10}, the $GdW$ corrections are
substantially reduced as compared to gas-phase PTCDA
\cite{footnote5}. 
As shown in Fig.~\ref{fig_Parameters}a, the QP corrections are 
below 0.5 eV for $z$\,=\,5\,\AA\, instead
of almost 2\,eV for the free molecule. Note, however, that as the
molecule is removed from the metallic substrate, its screening
becomes less metallic, and both $W$ and $\Delta^{GdW}$ grow again
with increasing $z$.
(ii) The $GdW$ correction of the LUMO changes sign 
as a function of its occupation.
If the state were empty (occupied) it would observe a $GdW$ correction 
of $\Delta^{GdW}_{n=0}$$>$0 ($\Delta^{GdW}_{n=1}$$<$0), as shown 
in Fig.~\ref{fig_Parameters}a.
At partial occupation 0$<$n$<$1 (through charge transfer from the metal)
we obtain an 
$n$-dependent QP shift $\Delta^{GdW}(n) = (1-n)\Delta^{GdW}_{n=0}
+ n\Delta^{GdW}_{n=1}$.
(iii) Since the QP correction $\Delta^{GdW}(n)$
shifts the whole LUMO spectrum $f(E)$ rigidly with respect to the
Fermi level, thereby changing its occupation by $\Delta n$, the
classical (screened) Coulomb energy changes by $2(\Delta n)U$ (the
factor of 2 results from spin degeneracy), which must be added to
$\Delta^{GdW}$, yielding at effective $GdW$ correction of
$\Delta^{GdW}_{\rm eff}(n) = (1-n)\Delta^{GdW}_{n=0} + 
n\Delta^{GdW}_{n=1} + 2(n-n_{\rm LDA})U$.
Since the
QP shift in turn determines $n=
\int_{-\infty}^{E_F} f^{GdW}(E) dE$ via $f^{GdW}(E)=f^{\rm
LDA}(E-\Delta^{GdW}_{\rm eff}(n))$, $n$ and $\Delta^{GdW}_{\rm eff}$
must be determined
self-consistently for each $z$ (Fig. \ref{fig_Parameters}a,c). 
At small $z$ we obtain a negative QP shift and a slight increase of $n$
(as compared to $n_{\rm LDA}$) while at large $z$ the QP shift 
approaches zero.
A key ingredient into our calculations is the intra-state interaction 
energy $U$. It is given (for each $z$) by
\begin{equation}
\label{eq_3} U = \int |\psi_{\rm LUMO}({\bf r})|^2 W({\bf r},{\bf
r}')
         |\psi_{\rm LUMO}({\bf r}')|^2 d^3r d^3r' \quad .
\end{equation}
Fig. \ref{fig_Parameters}b reveals that as a
consequence of efficient metallic screening the $U$ parameter is
rather small (below 1~eV) for $z$\,$<$\,8\,\AA, but increases for
larger $z$. For gas-phase PTCDA it amounts to 3.0\,eV.

Results for the effective QP shift $\Delta^{GdW}_{\rm eff}$,
the $U$ parameter, and the self-consistent occupation $n$
are displayed in Fig.~\ref{fig_Parameters}a-c. Although the QP
correction turns out to be small, there is a clear trend from
$\Delta^{GdW}_{\rm eff}=-0.17$\,eV for $z$\,=\,5\,\AA\,to
$\Delta^{GdW}_{\rm eff}$\,$\simeq$\,0\,eV for $z$\,=\,12\,\AA.
This leads to an increased slope of the LDA+$GdW$ QP peak position
for $z$ between 7 and 10 \AA\, (Fig. \ref{fig_LUMO}d).

We finally extend our LDA+$GdW$
calculation by including dynamical correlation in the NRG
approach. Using the calculated values of $U(z)$,
Eq.~(\ref{eq_3}), we are able to extract the bare LUMO level
positions $\epsilon_0(z)$ and coupling functions $\Gamma (E,z)$
from the LDA+$GdW$ spectra at each $z$, by equating
\begin{equation}
\label{eq_4} f^{GdW}(E,z)\equiv \frac{1}{\pi}\Im
\frac{1}{E-\epsilon_0(z)-nU(z)-\Gamma(E,z)}.
\end{equation}
Together with $U(z)$, the so obtained $\epsilon_0 (z)$, $\Gamma
(E,z)$ enter the NRG calculation
\cite{BullaCostiPruschke2008,PetersPruschkeAnders2006} which
yields the many-body LUMO spectra displayed in
Fig.~\ref{fig_LUMO}b-c \cite{footnote6}.

The $z$-dependent renormalized peak position of the LDA+$GdW$+NRG data 
(full squares) can now be compared to the experimental 
data as shown in Fig.~\ref{fig_LUMO}d.
As in experiment,
the calculation shows a fast-rising QP peak (slope 0.15\,eV/\AA\,
for $z$ between 8 and 10 \AA), which also becomes pinned at
$E_F$ as soon as it reaches the Fermi energy (at $z$ $\approx$ 10
\AA). In both experiment and LDA+$GdW$+NRG, the pinned resonance
is substantially sharper than the QP peak that moves up to $E_F$
(see Fig. \ref{fig_LUMO} c). 
Since this sharpening to 14
meV (experiment at 9.7\AA) or 15 meV (theory at 11 \AA) is
neither observed in LDA nor in LDA+$GdW$, we can ascribe it to a
dynamical correlation effect.

Note that for $z$\,$\lesssim$\,8\,\AA\, the LDA+$GdW$ and LDA+$GdW$+NRG
spectra are very similar (Fig.~\ref{fig_LUMO}b). This is to be
expected, since a nearly fully occupied state below $E_F$ shows
very weak dynamical correlation. In contrast, for
$z$\,$\gtrsim$\,9\,\AA\, the difference between LDA+$GdW$ and
LDA+$GdW$+NRG becomes dramatic. While LDA+$GdW$ still yields a
single LUMO peak (Fig.~\ref{fig_LUMO}c), 
the LDA+$GdW$+NRG spectrum 
shows a three-peak structure, with two single-particle side bands
(at --0.9 eV and +0.3 eV for $z$=11 \AA, separated by $\sim$$U$)
and a Kondo resonance at $E_F$ (Fig.~\ref{fig_LUMO}c).
The Kondo resonance is also observed in
experiment (bottom of Fig.~\ref{fig_LUMO}c and sharp horizontal
line in Fig.~\ref{fig_LUMO}d).

In Fig.~\ref{fig_Parameters}d the evolution of the Kondo
temperature $T_K(z)$ is plotted. $T_K$ has been determined from
the NRG calculation as the temperature at which 40 \% of the local
moment in the LUMO is screened (Wilson criterion). For
$z$\,$<$\,9\,\AA\, $T_K$ is a few thousand K, because in this
$z$-range $U$ and $\Gamma$ are of similar size, and hence there is
no clear separation between spin and charge fluctuation energy
scales. In other words, in this regime $T_K$ is equivalent to the
charge fluctuation scale $\Gamma$. Physically, this means that
screening of the small residual moment that is left by the total
occupancy of 1.8 electrons in the LUMO is accomplished by
uncorrelated charge fluctuations in and out of the level. At
$z$\,$=$\,12\,\AA\, $T_K$ has dropped to $T_K$\,$<$\,1\,K,
caused by the increase of the ratio $U/\Gamma$
which leads to the clear separation of spin and charge
energy scales. Accordingly, $T_K$ is now a Kondo temperature in
the proper sense, related to the energy scale of correlated
screening of a fixed moment in the LUMO by virtual spin
fluctuations.
In experiment, the Kondo
resonance can only be followed up to $z$\,$\approx$\,10\,\AA\
because of decreasing signal strength, but at 9.7~\AA\ the Kondo
resonance has a FWHM of 14 meV, in good agreement with a predicted
Kondo temperature of the order 100 K (Fig.~\ref{fig_Parameters}d).

One remaining difference between our LDA+$GdW$+NRG data and experiment
is the $z$-position at which the QP peak reaches $E_F$ and turns into
the resonance.
This could be due to the underestimation
of the Ag(111) work function within the underlying DFT-LDA ($\sim$4.6 eV
instead of the experimental value of 4.8 eV), which persists throughout
our calculations.
Taking this into account, the Fermi level in all theoretical spectra
should be lowered by 0.2 eV, i.e. all molecular QP features in
Fig. \ref{fig_LUMO} would be at higher energies by 0.2 eV.
The LDA+$GdW$+NRG peak would then reach the Fermi level at about
$z$ $\approx$ 8 \AA.

In conclusion, we have presented an approach to calculating
electronic spectra that systematically includes non-local and
dynamic correlation effects. 
It correctly predicts the mechanical gating of the tip/PTCDA/Ag(111) 
junction.
Because of its computational efficiency, this state-of-the-art 
approach to electronic spectra should be widely applicable.

We thank the Deutsche Forschungsgemeinschaft for financial support 
(Grants RO 1318/6-1, AN 275/6-2 and TA 244/5-2).


\begin{thebibliography}{}
\bibitem{footnote1}
Previous studies of PTCDA/Ag(111) showed that DFT employing the
LDA yields reliable structural data
in good agreement with available experiments \cite{Hauschild05,Kraft06}
and with more elaborate many-body total-energy
approaches \cite{Rohlfing08}, although many details of long-range
correlation are missing from the DFT-LDA.
We thus employ DFT-LDA throughout this work.
\bibitem{footnote2}
Choosing Ag as tip material is motivated by the experimental tip
preparation method, in which a tungsten tip is dipped into the Ag
substrate, picking up some silver material.
Tungsten as tip material leads to equivalent results as discussed
here (apart from somewhat stronger tip-molecule bond energy 
for tungsten).
\bibitem{footnote0}
The LDA spectrum of the LUMO in contact
with the substrate results from projection of all states of the
respective system ($|\psi_n\rangle$) onto $|\psi_{\rm LUMO}\rangle$,
i.e.
$f_{\rm LDA}(E) = \sum_n | \langle \psi_n | \psi_{\rm LUMO}\rangle |^2
\delta(E-E^{\rm LDA}_n)$ (projected density of states, PDOS).
The occupancy of the state is
$n = \int_{-\infty}^{E_F} f(E) dE$.
\bibitem{footnote5}
For PTCDA on Ag(111), our LDA+$GdW$ method involves a further
approximation, i.e. the evaluation of $G$ from molecular states only
(i.e. from a calculation without substrate and tip),
disregarding the hybridisation with substrate states.
This is justified by the fact that the molecular states are only
slightly shifted and broadened due to contact with the substrate.
\bibitem{footnote6}
Since the LDA calculation indicates a
weak influence of the tip onto the spectra, we also neglect the
voltage dependence of the spectral function which would be only
significant for a symmetric junction \cite{Anders2008,Temirov08}.
\end{thebibliography}
\end{document}